\documentclass[11pt,fleqn]{article}
\usepackage{graphicx}
\usepackage[a4paper,left=1.5cm,right=1.5cm]{geometry}
\usepackage{amsmath}
\usepackage{float}
\usepackage[latin1]{inputenc}
\usepackage{amsmath}
\usepackage{amsfonts}
\usepackage{amssymb}
\usepackage{float}
\floatstyle{boxed} 
\restylefloat{figure}

\newcommand{\be}{\begin{equation}}
\newcommand{\ee}{\end{equation}}
\newcommand{\bea}{\begin{eqnarray}}
\newcommand{\eea}{\end{eqnarray}}

\begin{document}
\title{New tensor interaction as the source of the observed CP asymmetry in $\tau \rightarrow K_{S}\pi\nu_{\tau}$.}
\author{Lobsang Dhargyal \\\\\ Institute of Mathematical Sciences, Chennai 600113
, India.}

\maketitle
\begin{abstract}

Babar collaboration has reported an intriguing opposite sign in the integrated decay rate asymmetry $A_{cp}(\tau \rightarrow K_{S}\pi\nu_{\tau})$ than that of SM prediction from the known K0 - $\bar{K0}$ mixing. Babar's result deviate from the SM prediction by about 2.7$\sigma$. If the result stands with higher precision in the future experiments, the observed sign anomaly in the $A_{cp}(\tau \rightarrow K_{S}\pi\nu_{\tau})$ can most likely come only from some new physics occurring possibly in both hadronic as well as leptonic sectors. In this work, we will give an improved analysis of our previous work where we have illustrated that, while non-standard scalar or vector/axial-vector interactions will not contribute to the observed asymmetry in the integrated decay rate, a new tensor interaction can explain the observed anomaly. Assuming the real part of the new tensorial coupling is negligible compare to its imaginary part and with $A_{cp}(\tau \rightarrow K_{S}\pi\nu_{\tau})$ and Br($\tau \rightarrow K_{S}\pi\nu_{\tau}$) as data points to fit the imaginary part of the NP coupling, we have been able to fit the result within 1$\sigma$ of the experimental values.

\end{abstract}

\section{Introduction}

The study of CP violation in tau decays has always been of much interest for beyond the Standard Model studies in the past two decades. In SM, the only source of CP violation is the one phase in the Kobayashi Maskawa (KM) matrix. While the Kobayashi Maskawa ansatz for CP violation within the Standard Model in the quark sector has been clearly verified by the plethora of data from the B factories, this is unable to account for the observed baryon asymmetry of the Universe. Hence, one needs to look for other sources of CP violation, including searches in the leptonic sector. Apart from the CP phases that may arise in the neutrino mixing matrix, the decays of the tau lepton may allow us to explore nonstandard CP-violating interactions. Various experimental groups have been involved in exploring CP violation in tau decays in the last decade or more. The BABAR collaboration for the first time reported a sign anomaly in the integrated decay rate asymmetry $A_{cp}(\tau \rightarrow K_{s} \pi \nu_{\tau})$ of
\be
A^{Exp}_{cp} = (-0.36 \pm 0.23 \pm 0.11)\%.\\
\ee
However for $\tau^{\pm} \rightarrow K_{s}^{0} \pi^{\pm} \nu_{\tau} \rightarrow [\pi\pi]^{0}_{K}\pi^{\pm} \nu_{\tau}$, Babar has predicted the SM integrated decay-rate asymmetry to be
\be
A^{SM}_{cp} = (0.33 \pm 0.01)\%.\\
\ee
Comparing the rate asymmetries for decays to neutral kaons of the taus with that of D mesons, Grossman and Noir have pointed out that since $\tau^{+}(\tau^{-})$ decays initially to a $K^{0}(\bar{K^{0}})$ whereas $D^{+}(D^{-})$ decays initially to $\bar{K^{0}}(K^{0})$, the time-integrated decay-rate CP asymmetry (arising from oscillations of the neutral kaons) of $\tau$ decays must have a sign opposite to that of D decays. The observation of a CP asymmetry in $\tau$ decays to $K_{s}$ having the same sign as that in D decays, and moreover of the same magnitude but opposite in sign to the SM expectation, implies that this asymmetry cannot be accounted for by the CP violation in $K^{0} \bar{K^{0}}$ mixing.

\begin{enumerate}

\item Naively one may expect that the simplest way to account for the observed anomaly would be to introduce a direct CP violation via a new CP violating charged scalar exchange. However, it turns out that the charged scalar type of exchange may contribute in the angular distributions, but its mixing with SM term in the integrated decay rate goes to zero.

\item Now the next candidate of NP would be a new CP violating charged vector exchange, but CP violation from vector type NP will be observable only if both vector current and axial vector currents contributes to the same final states. Since in two pseudo scalar meson final states only vector current can contribute due to parity conservation of strong interaction, vector type of NP can contribute in general to CP violation in three or more pseudo scalar meson final states but not in two pseudo scalar meson final states such as $K_{s}\pi$.

\item Now the only possibility left is tensor type of NP.

\end{enumerate}

In this presentation, we will give the results the materials contianed in the references \cite{our1} and \cite{our2}.

\section{Results}

With taking the approximation of $A^{k}_{pc}A^{\tau}_{cp} \approx 0$ we can express the Eqs(29,30) of \cite{our2} as:
\be
A_{cp}(\tau \rightarrow K_{s}\pi\nu_{\tau}) = A_{cp}^{K} + A_{cp}^{\tau}
\ee
and
\be
Br(\tau \rightarrow K_{s}\pi\nu_{\tau}) = \frac{(\Gamma^{\tau^{+}} + \Gamma^{\tau^{-}})}{2}\tau_{\tau} = (\Gamma_{SM} + \Gamma_{T})\tau_{\tau} 
\ee
where $A^{k}_{cp}$ is the known SM CPV from the $K-\bar{K}^{0}$ mixing, $\Gamma_{SM}$ is the SM decay rate corresponding to fitted form factors from Belle fits, $\Gamma_{T}$ is the decay rate due to purely tensor term and $\tau_{\tau}$ is the life time of $\tau$ lepton. From Eqs(9,10) and using $F^{a}_{T}$ from Eqs(7) the best fitted value of the complex parameter $Im(C^{\tau}_{T})$ to the two data points gives at $\chi^{2} \approx 4.5$ :
\be
Im(C^{\tau}_{T}) =  -0.071,
\ee
which gives
\be
Br(\tau \rightarrow K_{0}\pi\nu_{\tau})^{(Th)} = 2Br(\tau \rightarrow K_{s}\pi\nu_{\tau})^{(Th)} = (0.756 \pm 0.085)\%\\
\ee
and
\be
A^{\tau(Th)}_{cp} = (-0.703 \pm 0.54)\%\\
\ee
whereas the experimental values of these observables are given as 
\be
A^{(Exp-SM)}_{cp} = A^{\tau(Exp)}_{cp} - (A^{k}_{cp})^{SM} = (-0.69 \pm 0.26)\%,\\
\ee
and
\be
Br(\tau \rightarrow K_{0}\pi\nu_{\tau})^{(Exp)} = 2Br(\tau \rightarrow K_{s}\pi\nu_{\tau})^{(Exp)} = (0.84 \pm 0.04)\%.\\
\ee
Comparing Eqs(7,8) and Eqs(6,9) we see that the theoretical predicted values fit with the experimental values within 1$\sigma$.

\section*{Acknowledgements}

Author did part of this work with Nita Sinha and H. Zeen Devi, Institute of Mathematical Sciences.

\end{document}